# Being published successfully or getting arXived? The importance of social capital and interdisciplinary collaboration for getting printed in a high impact journal in Physics.

Oliver Wieczorek [a], Mark Wittek [b], Raphael H. Heiberger [c]

[a] Otto-Friedrich-University of Bamberg, Department of Sociology, especially Sociological Theory.
Address: Feldkrichenstraße 21, 96052 Bamberg, Germany.
E-mail: oliver.wieczorek@uni-bamberg.de

[b] University of Cologne, Institut für Soziologie und Sozialpsychologie.
Address: Universitätsstraße 24, 50923 Köln. Germany.
E-mail: wittek@wiso-uni-koeln.de

[c] University of Stuttgart, Institute of Social Sciences, Department of Computational Social Science.
Address: Seidenstraße 36, 70174 Stuttgart. Germany.
E-mail: raphael.heiberger@sowi.uni-stuttgart.de

## Abstract

The structure of collaboration is known to be of great importance for the success of scientific endeavors. In particular, various types of social capital employed in co-authored work and projects bridging disciplinary boundaries have attracted researchers' interest. Almost all previous studies, however, use samples with an inherent survivor bias, i.e., they focus on papers that have already been published. In contrast, our article examines the chances for getting a working paper published by using a unique dataset of 245,000 papers uploaded to arXiv. ArXiv is a popular preprint platform in Physics which allows us to construct a co-authorship network from which we can derive different types of social capital and interdisciplinary teamwork. To emphasize the "normal case" of community-specific standards of excellence, we assess publications in Physics' high impact journals as success. Utilizing multilevel event history models, our results reveal that already a moderate number of persistent collaborations spanning at least two years is the most important social antecedent of getting a manuscript published successfully. In contrast, inter- and subdisciplinary collaborations decrease the probability of publishing in an eminent journal in Physics, which can only partially be mitigated by scientists' social capital.

# 1. Introduction

Why do some research papers become highly rewarded articles, while most others disappear in a discipline's file drawer (Rosenthal, 1979)? Sociology of science, scientometrics, and network science suggest that considering the social antecedents of knowledge production can shed light on this question (Blau, 1994; Crane, 1972; Jansen et al., 2009; Kuhn, 2012; Latour, 1987; Merton, 1968; Newman, 2001). Investigating the link between scientific collaboration and the chances for publishing an influential paper is getting ever more important, since academia is an increasingly social matter carried out in teams of growing sizes (Wuchty et al., 2007). Against this backdrop, collaborations are seen more and more frequently as the basic requirement for doing science (Barabâsi et al., 2002; Brint, 2001; Leahey and Barringer, 2020; Leahey and Cain, 2013; Moody, 2004) and are, in general, considered to improve chances to publish papers (Abbasi et al., 2018; Abramo et al., 2013; Lee and Bozeman, 2005). Furthermore, collaborations are the conduits through which scientists can build and access resources—such as funding and equipment—necessary to conduct expensive research (Boardman and Ponomariov, 2014; Knorr-Cetina, 1999).

Besides the importance of collaborations, scholars emphasize that scientific communities develop distinct standards of knowledge production leading to the consolidation of paradigms and boundaries that distinguish a community from other (sub-)disciplines (Bourdieu, 1988; Fuchs, 2009; Kuhn, 2012; Wieczorek and Schwemmer 2020). Being part of such a community and knowing its scientific norms is essential for getting published and—by doing so—for getting acknowledged academically (Münch, 2014, pp. 13–20). At the same time, interdisciplinary collaboration is increasingly demanded by funding agencies (Leahey and Barringer, 2020) and is seen as a source of new research strands as well as of enhanced scientific impact (Biancani et al., 2018; Lynn, 2014; Shi et al., 2009).

Whereas previous research focused on either the role of collaborations (Lazega et al., 2006; Li et al., 2013; Pezzoni et al., 2012) or interdisciplinarity (Biancani et al., 2018; Leahey et al., 2017; Millar, 2013; Uzzi et al., 2013) for scientists' publication success, we study both factors along with their interplay. In addition, we contribute to the existing literature by applying an analytical strategy that allows us to distinguish between published and unpublished papers, while previous research was limited to data on successfully published work.

To arrive at a comprehensive view of the role of collaboration partners for academic success, we firstly investigate three types of a scientist's social capital: the sheer number of co-authorships (Lee and Bozeman, 2005), long-term collaborations (Burt, 2001; Cainelli et al., 2015; Dahlander and McFarland, 2013), and collaboration with highly connected scientists (Han, 2003). Secondly, we consider how interdisciplinary collaboration and thus the chance of bridging epistemic boundaries (Burt, 2004, 2002; Patulny and Svendsen, 2007, p. 37f.) is linked to successful submissions in prestigious journals (Abbasi et al., 2014). While various studies analyze the benefits of work with an interdisciplinary orientation (Biancani et al., 2018; Leahey et al., 2017; Millar, 2013), we distinguish between collaboration among scholars with different disciplinary backgrounds and collaboration among scholars located in different subdisciplines.

These types of collaboration are possibly aligned with distinct paradigms of knowledge production (see for example Au, 2018; Wieczorek and Schwemmer 2020). In this way, we analyze how various kinds of social capital and interdisciplinary collaboration are linked to successful publication. Thirdly, we investigate the impact of the interplay between different types of social capital and interdisciplinary collaboration. This approach reveals who can afford to use a high risk high reward (Uzzi et al., 2013) strategy, adding a further nuance to the comprehensive existing literature on interdisciplinary research.

To assess the association between social capital, interdisciplinarity, and scientific impact, we investigate one of the most successful endeavors in the history of science: modern Physics (Kragh, 2002). Often regarded as "big science" (e.g. Galison, 1992), modern Physics is characterized by huge projects with large numbers of collaborating physicists to facilitate scientific breakthroughs (Heinze et al., 2009; Wuchty et al., 2007), and to handle the massive amount of data produced, e.g. by particle colliders or space-telescopes (Bodnarczuk and Hoddeson, 2008). Both the necessity to collaborate and its importance for sectors outside academia make Physics an ideal subject to study the social antecedents of knowledge production.

ArXiv is the central pre-publication platform in Physics. It facilitates scholarly communication and allows us to trace whether or not a paper gets published in a high impact outlet. The rapid progress of the internet during the 1990s brought about a sheer explosion of the pre-refereed distribution of scientific papers resulting in a total number of downloaded papers exceeding 1.6 billion (arXiv, 2020). This renders arXiv not only an important preprint platform for physicists, but also for researchers conducting meta-analyses (Henneken et al., 2006).

We collected over 245,000 arXiv preprints, from which we derive co-author networks to assess the social capital, as well as the inter- and subdisciplinary resources employed in the collaboration efforts. Utilizing multilevel event history models our results reveal that persistent collaborations show the strongest association with academic success, whereas interdisciplinary collaboration on its own does not increase the chances for publishing a manuscript in an eminent Physics outlet. This is especially true for physicists with low amounts of social capital, whereas well-connected scholars profit from interdisciplinary collaboration if they have persistent collaborations at their disposal.

## 2. Theory and Hypotheses

Social capital is inscribed in social relations between actors (Coleman, 1990; Lin, 2002) and stems from the access of resources inhabited by groups or networks (Bourdieu, 1986). These resources include domain-specific knowledge, funding, scientific reputation, or technical equipment (e.g. the Large Hadron Collider of CERN). Social capital includes both direct and indirect access to such resources provided by academic peers and peers of peers. Furthermore, social capital relies on time and repeated efforts of actors to build trust and affection (Bourdieu, 1988, p. 87), thus leading to network closure and bonding dynamics (Burt, 2001, 2000). In turn, trust and affection lead to a group identity and shared routines accountable for competitive advantages, i.e., writing high-quality papers published in high impact outlets. Insofar, social capital covers not only the number of social relations and the resources therein, but also the elements ascribed to constructs such as bonding capital (Breiger, 2010; Burt, 2007), strong ties (Cainelli et al., 2015; Granovetter, 1983) and persistent ties (Dahlander and McFarland, 2013). Applied to our case, social capital embodied in different types of collaborations among physicists covers multiple conditions relevant for scientific success. We therefore concentrate on three varieties of social capital: the number of collaborations, persistent collaborations, and collaborations with influential actors with abundant social resources at their disposal.

We regard the first type of social capital as being associated with successful publication is the number of co-authors. Collaborations are the type of scientific relations most frequently studied by previous accounts (Barabâsi et al., 2002; Ferligoj et al., 2015; Heiberger and Riebling, 2015; Jansen et al., 2009; Kronegger et al., 2011; Leahey and Cain, 2013; Lee and Bozeman, 2005; Mason, 2020; Millar, 2013; Newman, 2001; Ponds et al., 2007; Rossier, 2020; Uzzi et al., 2013). These studies establish positive associations between the number of co-authors and the number of papers published in scholarly outlets across many disciplines. Following our definition of social capital, being able to establish a large number of research collaborations increases the access to domain-specific knowledge, resources and information. These, in turn, are expected to increase the chances to publish a manuscript in an eminent outlet. For these reasons, we hypothesize:

H1.1. The likelihood of getting an arXiv preprint published in a high impact journal in Physics increases along with the number of a physicists' collaborations.

Besides the number of network ties, previous research suggests that the stability of collaboration has an impact on publication success. In line with Bourdieu (1985), we interpret social capital as the access to resources of a group or class of actors, which is granted by membership. However, group membership entails the necessity to invest time to establish trust, affective relations and a shared group identity (Bourdieu, 1987, 1985; Frederiksen, 2014). Those things, as well as an efficient division of labor, need time to develop and are more likely to appear among long time members of a research team (Boardman and Ponomariov, 2014; Cainelli et al., 2015; D'ippolito and Rüling, 2019; Krackhardt, 1992). It makes a decisive difference whether collaboration takes place between people who have known each other for a long time or whether they collaborate for the first time (Dahlander and McFarland, 2013). As Brint (2001, pp. 399–400) suggests, it is crucial to have "a small circle of people whom [the researchers] talk to regularly about their work". In addition, the findings of Leahy and Cain (2013, pp. 936–941) highlight that being part of a stable collaboration network is essential for succeeding in academia. We therefore expect:

H1.2. Persistent collaborations increase a physicist's chances of getting an arXiv preprint published in a high impact physics journal.

Along with the number of connections and access to group resources, social capital also implies with whom a scholar maintains these connections (Bourdieu, 2013). Status 'leaks through' social relationships (Podolny, 2010, 2001), meaning that a physicist's possibilities to publish papers prominently and to maintain a career in academia are influenced by the status of her or his associates. We assume that social capital partially follows a "prestige principle" (Han, 2003) which plays an additional role in the formation of individual scientists' collaborations (Bozeman and Corley, 2004). Highly reputed scholars have the ability to shape the epistemic norms of the field and they provide access to equipment and funding necessary to conduct research, thus collaborating with them supposedly increases the chances of getting a preprint published in a high impact journal.

H1.3. The likelihood of getting an arXiv preprint published in a high impact journal in Physics increases along with the number of high-status collaborators.

We assume that different dimensions of social capital complement each other when it comes to the chances of publishing in high impact journals. As, e.g., Podolny (2010) points out, affiliations with high-status actors heighten an actor's status, which in turn increases his or her prominence as a collaboration partner. In our case, connections to prolific scientists should lead to more co-authorships since individuals profit from such relations and become more attractive for others. To take such interdependencies into account, we include interaction terms between different types of social capital:

H1.4. The likelihood of getting an arXiv preprint published in a high impact Physics journal increases as a physicist has multiple types of social capital at his or her disposal.

Besides different types of social capital, interdisciplinary collaboration is increasingly seen as a requirement for high impact and innovative research (Biancani et al., 2018; Leahey et al., 2017; Millar, 2013; Rawlings et al., 2015; Uzzi et al., 2013). Actors that are in a position to connect two domains of knowledge (e.g. Biology and Computer Science) are said to gain competitive advantages stemming from the different domains and are coined 'boundary spanners' (Aldrich and Herker, 1977; Chau et al., 2017; Kaplan et al., 2017). They assume the role of 'information brokers' (Burt, 2004) and combine the knowledge of different network domains to find more creative solutions (Hansen, 1999).

While scholars find potential positive effects of interdisciplinary work (Larivière and Gingras, 2010; Millar, 2013), there is little empirical work available investigating whether interdisciplinary collaboration fosters or impedes individuals' academic success (for exceptions, see Hackett and Rhoten, 2009; Leahey et al., 2017). Studies focusing on the importance of paradigms and associated disciplinary standards suggest that leaving the boundaries of a discipline might be a risky strategy, since publication outlets are tied to paradigms (Grothe-Hammer and Kohl, 2020; Wieczorek and Schwemmer 2020). The latter is backed by evidence from other fields of cultural production—such as the Broadway musical industry. Studies suggest that there is a trade-off between closure and ties spanning separated areas of a wider collaboration network and that a balance between these two components offers benefits in terms of creative and financial performance (Uzzi and Spiro, 2005). Furthermore, some audiences of cultural products might be more reluctant than others towards attempts of cultural producers to span established boundaries (Goldberg et al, 2016).

In line with literature on potential negative effects of bridging and findings provided by Uzzi et al. (2013), we expect crossing interdisciplinary boundaries to be a risky strategy for two reasons. Firstly, mixing disciplinary standards of knowledge production is likely to lead to irritated reviewers who are most familiar with their own (sub)disciplines' norms of how to conduct high-quality research. Secondly, problems in coordinating interdisciplinary research efforts may arise due to the different socialization of the participants (see Gardner et al., 2014). Because of the mixed evidence, we probe deeper into the negative link between interdisciplinarity and academic success and hypothesize:

H2.1. If a physicist's collaboration spans at least two academic disciplines, it decreases her or his chances of getting an arXiv preprint published in a high impact Physics journal.

Since Physics is one of the largest and most differentiated scientific fields, we further differentiate between interdisciplinarity and collaborations spanning different subdisciplines in Physics (e.g. Nuclear Physics and High Energy Physics). It is plausible to assume that such efforts also span intellectual boundaries and respective epistemological cultures at the intersection of subdisciplines (Heidler, 2011; Leahey et al., 2017). We therefore hypothesize:

H2.2. If a Physicist's collaboration project is spanning multiple fields within Physics, it decreases the chances of getting an arXiv preprint published in a high impact Physics journal.

In contrast to our general assumption of a negative link between the bridging of interdisciplinary and subdisciplinary domains and high impact publications, there is also evidence of the positive effects of interdisciplinarity on, e.g., scholars' publication output (Abbasi et al., 2018; Abramo et al., 2013; Leahey and Barringer, 2020). From a theoretical perspective, one could argue that interdisciplinary collaboration should lead to more creative and recognized research because it bridges disconnected communities and thereby profits from a flow of non-redundant information (Burt, 2004; Shi et al., 2009). Yet, as Burt (2005) argues, brokerage in itself might fail to exhibit its full potential if actors lack close ties that are necessary to distribute information accessible through a brokerage position. He argues that a combination of network closure and brokerage provides maximum gains in work team performance, for instance, since brokerage can counteract potential negative aspects of too close-knitted networks and vice versa (Burt, 2005, pp. 131–146). Therefore, we assume that physicists need a stock of social capital to overcome the field-inherent barriers to inter- and subdisciplinary collaboration and reap the fruits of these high risk strategies (Martín-Alcázar et al., 2020; Uzzi et al., 2013). In line with Nahapiet and Ghoshal (1998), we therefore expect all types of social capital to help reverse the potentially negative effect of spanning different epistemic cultures:

H3.1. If a physicist's collaboration project spans multiple scientific fields and has a high amount of social capital, this increases the chances of an arXiv preprint to be published in a high impact Physics journal.

H3.2. If a physicist's collaboration project spans different subfields of physics and has a high amount of social capital, this increases the chances of an arXiv preprint of getting published in a high impact Physics journal.

## 3. Data and measures

ArXiv was established in 1991 and built to serve about 100 submissions per year, assumed to stem from a then small subfield of High-Energy Physics. The editorial directors of both major physical associations in the US, the American Physical Society and the Institute of Physics Publishing, quickly appreciated the system's benefit as an actual archive and, even more so, as a way of global distribution of ideas. The rapid progress of the internet led to a sheer explosion of the platform and made it very popular in Physics and related disciplines like Mathematics or Computer Science (Ginsparg, 2011).

We retrieved the arXiv data from the official Application Programming Interface (API) using Python and downloaded all articles related to Physics. The API contains several instances of meta-information from each paper including upload date, date of publication, publication outlet, if applicable, and author names. Using author information linked to arXiv preprints, we construct a network of co-authorship relations to derive our measures of social capital, subdisciplinary collaboration and interdisciplinary collaboration.

Each author represents a node in the network derived from the arXiv preprints. Edges between authors were assigned if they co-authored a paper. Based on the assumption that research teams are characterized by a division of labor and each collaborator writes only parts of the article, we weighted the strength of the potential collaboration by the numbers of co-authors per article. By doing so, we assumed that the physicists contributed equally to the manuscript. For instance, if two physicists co-authored a paper, each gets a weighted degree value of 0.5.

After cleaning procedures (see appendix A), our data set comprises 301,989 papers uploaded to arXiv between 2001 and 2014. Albeit our data starts in 2001, we decided to exclude data on research collaborations from the year 2001 in order to calculate the effect of persistent collaboration.

The final sample concentrates on the years until 2011 to allow papers three years to get published. It contains 245,432 arXiv preprints issued by 108,348 different authors. The authors accomplish 88,004 publications in high impact journals (35.86%), 62,676 were published in low impact journals (25.54%), while 94,752 (38.61%) papers were not published at all by the end of 2014. Figures 1 and 2 give an overview of the collaboration structure of the field of Physics in 2002 and 2011 as extracted from the arXiv data.

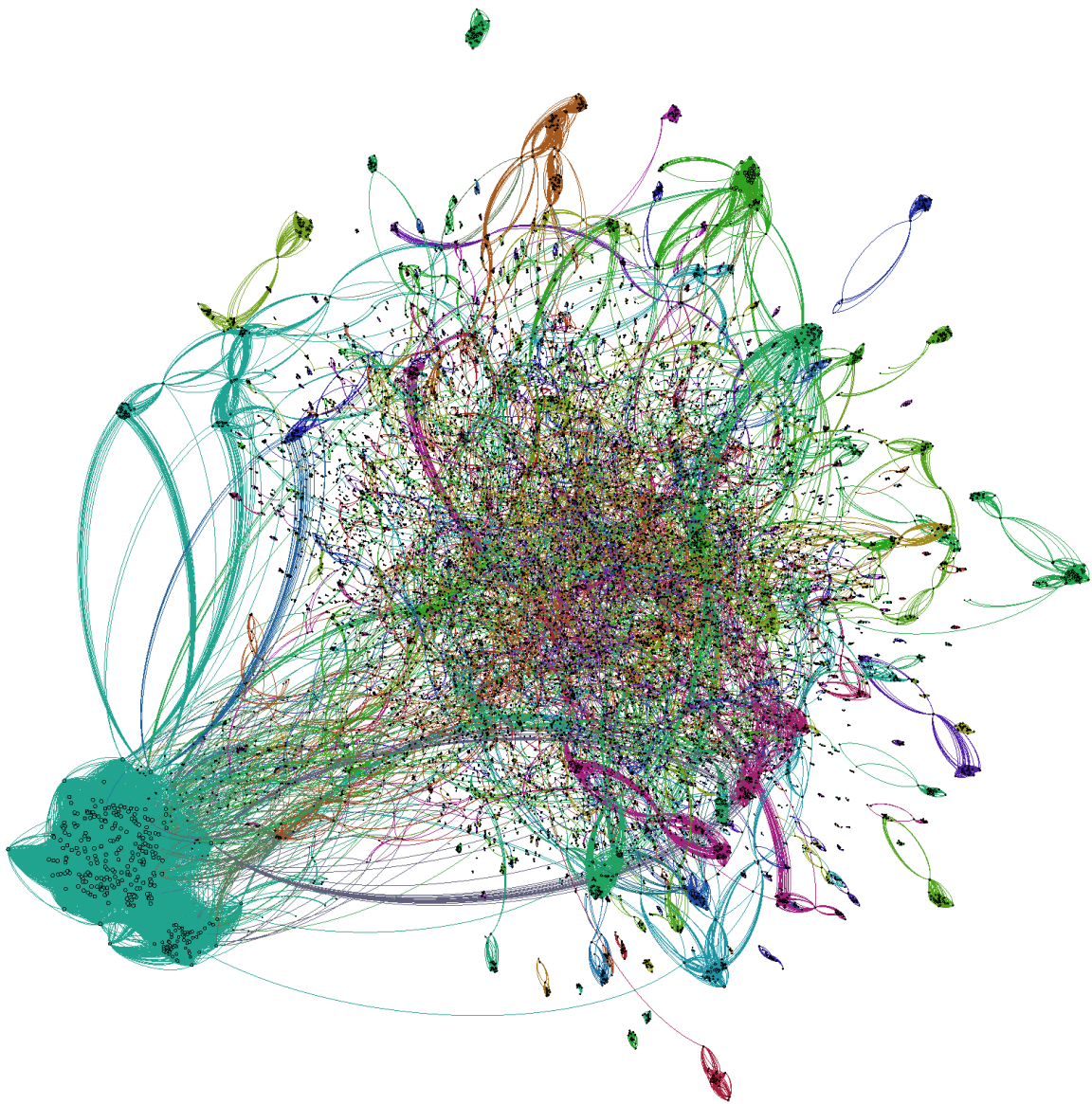

**Figure 1.** Structure of the collaboration network of Physicists on arXiv in 2002 consisting of 20,464 authors and 86,237 collaborations.

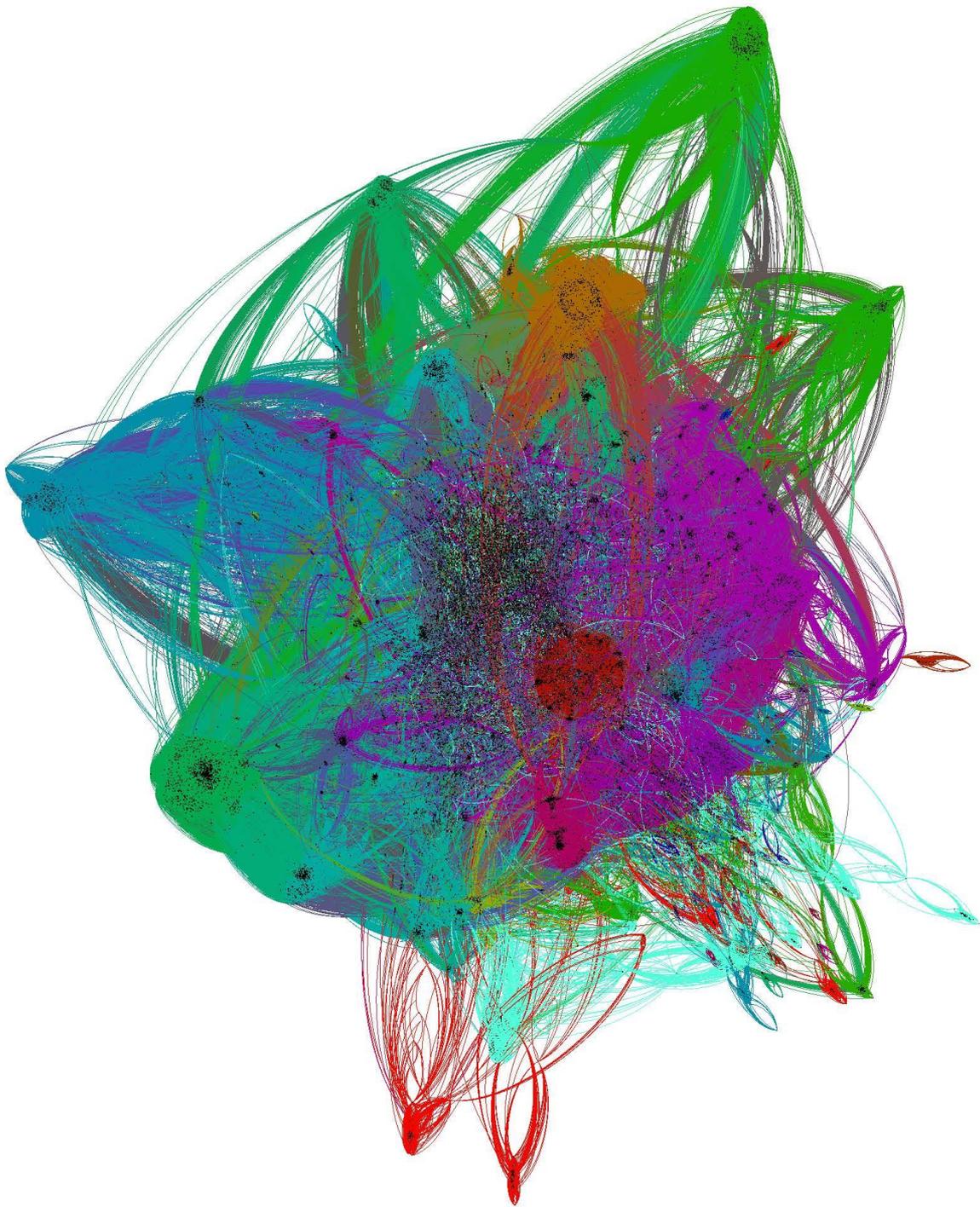

**Figure 2.** Structure of the collaboration network of Physicists on arXiv in 2011 consisting of 64,785 authors with over 1 million collaborations.

To construct the dependent variable, we first of all distinguish between unpublished and published arXiv preprints. Furthermore, we assume differences in quality and reputation between high impact journals and non-high impact journals in Physics. We declare the ten journals with the highest five-year impact factor of each Physics subdiscipline as high impact journals as they are the most selective, most visible and thus prestigious outlets in each subdiscipline (cf. details appendix B). Therefore, getting an arXiv preprint published in one of these journals constitutes our dependent variable.

Turning to the first dimension of social capital, we operationalize the number of co-authorships as the weighted degree of each author to test H1.1. Due to the extremely right skewed distribution of social capital, we used the percentiles of the distribution for correct parameter estimation.

Persistent collaborations represent the time consuming dimension of trust of social capital (Bourdieu, 1986). We define persistent collaborations as collaborations spanning at least two successive years to test H1.2. Since a large number of authors issued only one publication and the percentile approach produced only a near-dichotomous distribution, we decided to treat persistent ties as dichotomous variable.

Social capital accumulated by prominent Physicists is measured by their eigenvector centrality, as established scholars tend to collaborate with their likes (Bonacich, 1972; Burris, 2004). This depicts the aspect of social capital associated with leakage of prestige (Podolny, 2010, 2001) and status (Burris, 2004) and the access to resources present in groups and network structures. We decided to use percentiles for the calculation of the eigenvector centrality for the same reasons as for weighted degree. An overview of the descriptive statistics of the three network-related measures and further descriptive statistics on the DV are provided in Appendix C. To test H2.1 and H2.2, we use interdisciplinary and subdisciplinary collaboration to account for the capacities of physicists to bridge two or more paradigmatic domains.[1] To do so, we construct a categorical variable assigning each paper a status of either being located in a single domain of physics (interdisciplinary = 0), spanning between different subdisciplines of Physics according to Web of Science categorization (e.g. Nuclear physics and High Energy physics with interdisciplinary = 1), or spanning between disciplines such as Nuclear Physics and Engineering (interdisciplinary = 2). Table 2 shows the absolute and relative frequencies of the three types of collaboration.To investigate the interplay between interdisciplinary collaboration and social capital as well as their impact on getting an arXiv preprint successfully published, we calculated interaction effects between the percentiles of weighted degree centrality, eigenvector centrality, and our measure for inter- and subdisciplinarity to test H3.1 and H3.2.

## 4. Method

To measure how different types of social capital, inter- and subdisciplinary collaboration, and their interaction are associated with publication success in Physics, we apply a two-level event history analysis (Steele, 2008). The multi-level approach is necessary because every arXiv-preprint is nested within an author. We decided to apply a parametric, repeated event model to investigate the chances of getting a manuscript published (De Nooy, 2011; Windzio, 2006, p. 176). We chose to do so, since event history models account for changing chances to get a paper published over time in a high impact Physics journal. To this end, they use the conditional probability for event occurrence given that the event did not occur before. In our case the models take into account how the chances to publish in a high impact journal a year after submission change on condition that the paper was not published the year before. Thereby, we also account for the change in the numbers of preprints available for publication in high impact Physics journals over time.

---

[1] We decided to exclude Burt's constraint measure due to high levels of collinearity with the variables measuring persistent ties and inter- and subdisciplinary collaboration.

In this sense, an event occurs, if a manuscript is published in a high impact physics journal, whereas the contributions of the arXiv-preprints constitute the population at risk for being published (see Singer and Willett, 2003, p. 329). Every preprint uploaded between 2002 and 2011 is right censored (Scott and Kennedy, 2005, pp. 418–419), if it was not published in a high impact physics journal by the end of 2014. Since the exact upload time and date of acceptance are rarely included in our data, we decided to use years as time scale. If a preprint was published in a high impact physics journal in the same year, it would be published at t=1, indicating the first possible time to get printed. Articles published in 2002 could therefore be at risk for 12 years (t=13) and getting published at last after 11 years (t=12). Therefore, preprints uploaded at a later time (e.g. 2010) have higher chances of being right censored and are assumed to have no influence on the chances for publication of a preprint issued earlier (e.g. 2002). To take into account that papers filed for publication toward the end of the year have less chances of being published in that very year, we only consider articles uploaded until the 1st of September of each year for establishing values for the year in question.

Since every arXiv preprint has a distinct chance of getting published in a high impact Physics journal and is "nested" within a physicist, we assume a random intercept following normal distribution with a mean of zero. We do so to account for unobserved differences between physicists and to calculate the impact of social capital, interdisciplinary collaboration, collaboration among subdisciplines as well as their interaction on getting a paper published in a high impact physics journal.

The chances for an arXiv preprint to get published in a high impact physics journal and thus for physicists to successfully publish a manuscript is conveyed by the following hazard function which depicts our multilevel event history model:

$$h_i(t) = e^{\alpha(t)+\beta x_i(t)+u_i} \text{ with } \boldsymbol{u}_i \sim N(0, \sigma_u^2)$$

Here, $h_i(t)$ represents the hazard function for each arXiv preprint to get published between the beginning of 2002 and the end of 2014. It calculates the odds-ratios for getting a manuscript published in a high impact Physics journal, provided that it was not published the year before. The constant $\boldsymbol{\alpha}(t)$ represents the baseline chance of a manuscript for getting published in a high impact physics journal in each year, provided that it was not published the year before. $\boldsymbol{\beta x}_i(t)$ contains the effect sizes of a variable vector for each year. The variable vector **x** includes the measures of social capital (percentiles of weighted degree and eigenvector centrality, dichotomized persistent ties), the categorized measure for interdisciplinary collaboration and collaboration between different physics subdisciplines as well as their interactions.

Finally, $\boldsymbol{u}_i$ accounts for the fact that each physicist $\boldsymbol{u}$ is able to author different papers i. Since we aim to model the net-effect of our independent variable, $\boldsymbol{u}_i$ represents a random intercept and accounts for a fluctuation in publication chances "within" a physicist. Since $\boldsymbol{u}_i$ performs a normalization to account for fluctuations of publication chances within physicists, it follows a normal distribution with a mean of zero and publication chances vary by a standard deviation for each scientist in our data ($\boldsymbol{u}_i \sim N(0, \sigma_u^2)$).

Since we apply a parametric multilevel event history model, we test what type of distribution best fits our models (Blossfeld et al., 2019). To do so, we conducted a likelihood ratio test on our nullmodel, which suggested a Weibull-distribution (Table 1). The mean duration of a preprint before publication in a high impact journal is 4.47 years, with a medium of 5 years at risk, with a minimum of less than one year at risk and a maximum of 12 years at risk. This applies to the right censored preprints uploaded in 2002 and censored at the end of 2014. The number of preprints per scholar ranges from 1 to 13.

We use odds ratios to enhance the interpretability of our reported effects and use average marginal effects and t-values to compare the effect sizes of social capital, the categories of interdisciplinarity and their interactions. AIC and log-likelihood are used to compare the goodness of fit for each model.

**Table 1.** Log Likelihood Test of different parametrical distributions.

| Exponential Distribution | Gamma Distribution | Weibull Distribution | Lognormal Distribution |
|---|---|---|---|
| -321,747.12 | -292,500.04 | -165,235.26 | -297,701.47 |

As additional robustness checks, we calculated the full model with getting a preprint published in any journal and additionally applied a mixed effects panel regression (see Tables D1 and D2 in Appendix D).[2]

## 5. Results

In line with our theoretical expectations on the first dimension of social capital, we report a positive association between the weighted number of co-authorships and the likelihood to publish in a high impact journal (H1.1). The value of 1.019 indicates a 1.9% higher chance to get an arXiv preprint published in a high impact journal for each higher percentile of weighted degree. According to model 1, the total chances of getting a paper published in a high impact physics journal is 19.78% higher for physicists in the top percentile compared to Physicists in the lowest percentile of social capital associated with weighted degree.

---

[2] The main effects remain robust. We find some changes on the effect of degree and eigenvector centrality.

Persistent ties have the strongest effect in our models (H1.2). Having at least one persistent tie increases researchers' chances to publish in a high impact outlet by 23.66%. Compared to weighted degree, the effect of having at least one persistent collaboration is much stronger (t = 27.57 compared to t = 13.32), thus confirming the positive effect that relatively long-lasting collaborations have on creating successful publications.

According to model 1, we find, surprisingly, a negative link between having well-connected co-authors and publication success and therefore have to reject hypothesis H1.3. For every additional percentile, the chances of getting a paper published in a high impact journal diminishes by 4.945% (t = -36.32), resulting in 49.95% lower chances for physicists attributed to the top percentile compared to the lowest percentile.

Model 2 adds interaction terms between dichotomized persistent ties and weighted degree- as well as eigenvector centrality and renders the effects of social capital more complex than suggested by model 1. Now, the model implies a rejection of H1.1 and partial corroboration of H1.3. In the first case, the coefficient of degree centrality without persistent collaborations drops to 0.9387 (t = -10.92), indicating that if a physicist does not have access to persistent collaborations, the chances for getting an arXiv preprint successfully published drop by 6.13% for each additional percentile. Nevertheless, physicists with at least one persistent collaboration have a higher chance of getting a manuscript published successfully by 0.78% for each additional percentile.

**Table 2:** Exponentiated coefficients; *t* statistics in parentheses. Percentiles are used to calculate the effects of degree centrality and eigenvector centrality on the dependent variable.
* *p* < 0.05, ** *p* < 0.01, *** *p* < 0.001

| Dependent variable: Chance to publish an arXiv preprint in a high impact physics journal | Model 0 | Model 1 | Model 2 | Model 3 | Model 4 |
|---|---|---|---|---|---|
| Degree Centrality | | 1.0198*** | 0.9387*** | 0.9279*** | 0.9189*** |
| | | (13.32) | (-10.92) | (-12.96) | (-14.19) |
| Eigenvector Centrality | | 0.9506*** | 1.4121*** | 1.3923*** | 1.3859*** |
| | | (-36.32) | (55.79) | (53.54) | (51.52) |
| Persistent Ties: yes | | 1.2366*** | 1.2315*** | 1.2303*** | 1.1727*** |
| | | (27.57) | (26.97) | (26.77) | (15.75) |
| Persistent collaborations: yes * Eigenvector Centrality Categorized | | | 0.9657*** | 0.9670*** | 0.9668*** |
| | | | (-68.55) | (-65.83) | (-66.07) |
| Persistent Ties: yes * Degree Centrality | | | 1.0078*** | 1.0087*** | 1.0088*** |
| | | | (15.13) | (16.84) | (17.14) |
| Subdisciplinary Collaboration | | | | 0.6489*** | 0.3828*** |
| | | | | (-43.96) | (-37.63) |
| Interdisciplinary Collaboration | | | | 0.8657*** | 0.9232** |
| | | | | (-14.52) | (-3.22) |
| Subdisciplinary Collaboration * Degree Centrality | | | | | 1.0347*** |
| | | | | | (9.88) |
| Subdisciplinary Collaboration Persistent Ties: yes | | | | | 1.1101*** |
| | | | | | (5.45) |
| Subdisciplinary Collaboration * Eigenvector Centrality | | | | | 1.0498*** |
| | | | | | (13.29) |
| Interdisciplinary Collaboration * Degree Centrality | | | | | 1.0029 |
| | | | | | (0.83) |
| Interdisciplinary Collaboration * Persistent Ties: yes | | | | | 1.1204*** |
| | | | | | (5.90) |
| Interdisciplinary Collaboration * Eigenvector Centrality | | | | | 0.9807*** |
| | | | | | (-5.36) |
| *N* | 245432 | 245432 | 245432 | 245432 | 245432 |
| AIC | 324305 | 322510.2 | 318134.6 | 316132.5 | 315470.3 |
| Log lik. | -162149.5 | -161249.1 | -159059.3 | -158056.2 | -157719.2 |

However, if we introduce an interaction between This effect is reversed for eigenvector and persistent collaboration in model 2, we find that eigenvector centrality is strongly and positively significant if a Physicist *does not have* a single persistent collaboration at her or his disposal. In fact, for each additional percentile of eigenvector centrality, the chances for getting a preprint published successfully increase by 42.24% (t = 55.79). The opposite is true for high values of eigenvector centrality if we account for the interaction term with persistent collaborations. In this case, the chances to get a manuscript successfully published drop by 3.43% for each additional percentile. Taken together, model 2 falsifies H1.4 and shows that inheriting different types of social capital does not necessarily lead to higher chances to get an arXiv preprint published in a high impact Physics journal.

Model 3 adds interdisciplinary collaboration and subdisciplinary collaboration. Contrary to the notions that interdisciplinary collaboration is a panacea for scientific success (Biancani et al., 2018; Millar, 2013; Rawlings et al., 2015; Uzzi et al., 2013), we found neither interdisciplinary nor subdisciplinary collaboration to raise the chances of an arXiv preprint for publication in an eminent physics journal. On the contrary, chances decrease by 35.11% (t = -43.96) in the case of subdisciplinary collaboration among physicists, and by 13.44% (t = -14.52) for interdisciplinary collaboration between physicists and scholars belonging to other disciplines. Model 3 therefore corroborates H2.1 and H2.2.

In model 4, we include interaction terms between all three forms of social capital and interdisciplinarity as well as subdisciplinary collaboration. Beginning with the effect of different types of social capital and subdisciplinary and interdisciplinary collaboration without interaction terms, we witness their effects to remain stable. However, the interaction terms unfold a different and more nuanced picture of the effects of collaboration and interdisciplinarity on publication success in physics. We witness a drop in chances for an arXiv preprint to get successfully published from 35.11% (t = - 43.96) in model 3 to 61.73% (t = 37.63) in model 4, whereas the chances of getting an arXiv preprint published in a high impact physics journal slightly increase from – 13.44% (t = -14.52) to – 7.67% (t = 3.22). Especially the decrease in t-values between model 3 and model 4 indicates that the coefficient of interdisciplinary collaboration is instable and thus insufficient to explain differences in the chances to get an arXiv paper published in a high impact physics journal alone.

The interaction terms between different types of social capital, and subdisciplinary and interdisciplinary collaborations reveal two things. Firstly, they provide additional support for the relevance of persistent collaborations for getting an arXiv preprint published successfully (H.1.2) and the mechanism of trust and familiarity (Bourdieu, 1985; Martín-Alcázar et al., 2020) for gaining competitive advantage in academia (Munoz-Najar Galvez et al., 2019). Having at least one persistent collaboration dampens the negative coefficient of team members belonging to different epistemic cultures by 11.01% (t = 5.45) for subdisciplinary collaboration and by 12.04% (t = 5.90) for interdisciplinary collaboration. In fact, the negative association between interdisciplinary collaboration and academic success is cancelled out and partially falsifies H2.1, if persistent collaborations are present within the research team.

Secondly, similar effects of the interaction between social capital and subdisciplinary collaboration on publication probability appear only in the interaction with weighted degree (1.0347, t = 9.88) and eigenvector centrality (1.0497, t = 13.29). This finding corroborates H3.1 and shows, that spanning disciplinary boundaries while having high amounts of social capital increases the chances to publish an arXiv preprint in an eminent journal. In the case of an interaction between subdisciplinary collaboration and degree centrality, no significant association is present. At the same time, the interaction between eigenvector centrality and interdisciplinary collaboration is 0.9807, indicating that for each additional percentile, it gets more unlikely by 1.93% for an arXiv preprint to get published in a high impact physics journal provided that it was not published the year before. In line with the coefficients provided in model 4, our findings falsify H3.2.

To underline our interpretation, we depict the interaction between persistent collaborations and the other two types of social capital as average marginal effects to clarify the otherwise hard-to-interpret interaction effects (Mood 2010). Figure 3 reveals a "sweet spot" of collaboration for eigenvector centrality for the 5th percentile and an inverse effect for weighted degree centrality. Therefore, it seems to be the best strategy for getting an arXiv preprint published in a high impact physics journal if one has persistent ties, a high number of collaborations and a *medium* number of eminent colleagues at one's disposal for publication. Too many relations with eminent colleagues instead hinder those chances.

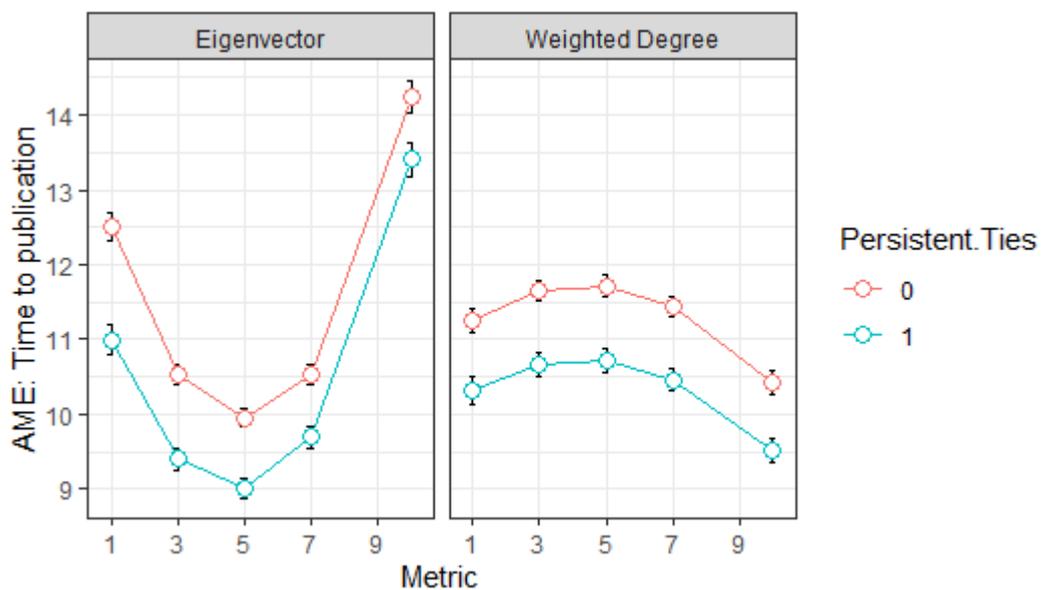

**Figure 3:** Average marginal effect predictions for getting an arXiv preprint published in a high impact physics journal. The Y-axis denotes time until publication, whereas the X-axis depicts the percentiles of the centrality values.

## 6. Discussion and conclusion

Our paper sought to answer the questions of whether 1) high amounts of social capital and 2) interdisciplinary or subdisciplinary collaboration make it more likely to get a manuscript published in a high impact journal. We avoid the survivor bias inherent in many previous studies by using a large number of manuscripts uploaded as preprints on arXiv, which allowed us to track whether or not they were published in high impact journals, low impact journals or not at all. We focus on Physics with its variety of subdisciplines and large impact on science's progress in general.

We differentiated social capital into three types: The number of collaborations (weighted degree), persistent collaborations built upon trust and familiarity (persistent collaborations) and the eminence of research collaborators (eigenvector centrality). Additionally, we tested whether interdisciplinary collaboration increases the chances for getting a manuscript published in a high impact journal in physics. We also investigated whether collaboration between physicists belonging to different subdisciplines bears negative consequences for getting a manuscript published in a high impact journal. We applied a multilevel event history model to account for the fact that every physicist uploaded at least one manuscript on arXiv.org and for a change in the probability to get a preprint published in a high impact journal over time. By doing so, our analytical strategy allows us to look at the successful papers and their authors and to consider failure in terms of papers that are not published in high impact physics journals or not published at all.

Our main findings emphasize the importance of persistent collaboration for academic success and are consistent with the findings of other case studies conducted by Leahey and Cain (2013), as well as Dahlander and McFarland (2013), Godechot (2016) and Ylijoki (2013). Furthermore, we show that a high number of well-connected co-authors tend to be negatively associated with academic success, which means that physicists with a large, but unselective number of collaborations tend to be less successful in getting their arXiv preprints published prominently in high impact journals. This can be ascribed to less familiarity among scholars and accordingly less effective sharing of information in line with Bourdieu (1985), lower levels of bonding social capital (Burt, 2001, 2000), or hints at less innovative normal sciences (Kuhn, 2012). Turning to the interaction between eigenvector centrality (as proxy for status, Burris 2004) and persistent collaboration, we unveil that stable and positive effects of status without any persistent collaboration might be linked to eminent physicists. These have the chance to attract constantly new, yet short-lived collaborations. In other words, eminent physicists might have the chance to choose their collaborators and to scrutinize the strength of weak ties (Granovetter, 1983).

The effects of interdisciplinary boundary spanning and being an information broker (Burt, 2004) often appears as panacea for academic success (Boardman and Ponomariov, 2007; Etzkowitz and Viale, 2010; Gaffikin and Perry, 2009; Lynn, 2014; Sá, 2008). However, our findings support a more critical view on this topic in line with Boden et al. (2011) or others who consider interdisciplinary research a "high risk, high reward" strategy (e.g., Uzzi et al., 2013). In all models, disciplinary diverse collaborations have a strong negative association with success.

Regarding the association between publication success and inter(sub)disciplinary collaboration among physicists our findings indicate alignments of collaboration projects to different epistemological communities of physicists. In line with Schwemmer and Wieczorek (2020) and Turner (2016), this alignment hinders scholars to publish their research in high impact journals due to the additional linkage between epistemic cultures and publication outlets. Different epistemological stances must be unified or translated. This procedure might be held accountable for a loss in analytical clarity and loss of originality as suggested by Münch (2014, pp 44–53).

Nevertheless, our findings are limited by a number of unmeasured characteristics of the papers and researchers being studied. On the one hand, institutional affiliations are not reported resiliently in the arXiv data. Only 7 percent of all authors specify their home institution, which would have meant a dramatic loss in sample size. The sheer number of cases and their longitudinal composition prevented us from merging this information manually. In this sense, one resulting potential bias relies on the fact that central positions of physicists in collaboration networks are mainly occupied by scholars located in the United States as Marginson (2006), Münch (2014) or Heiberger and Riebling (2015) suggest.

A second limitation is the omission of demographic attributes. We have no possibility to control, for instance, gender, race, etc. of the Physicists publishing on arXiv. One potential solution could be to draw up a smaller randomized sample that could be investigated for those individual covariates. From a data point of view this problem would be easier to handle than the first limitation, since those attributes do not change over time (in contrast to institutional affiliations). We try to account for that shortcoming by using a panel regression (appendix D).

A third limitation is rooted in the fact that arXiv is by design mostly relevant for Physicists. We are therefore unable to control whether our results are meaning ful for other disciplines or whether we discovered network effects specific to the community of Physicists. This could be a worthwhile direction for future research, since disciplines do have specific publication opportunities and restraints (Jansen et al., 2009; Whitley, 2000).

The fourth limitation is, that we are not in a position to make any strong claims about causality. We are unable to disentangle, whether the observed patterns are genuinely driven by co-authorship relation s, e.g. because they facilitate the transfer of field-specific knowledge, or whether other underlying factors such as being at the same laboratory (Silva et al., 2019), positive affection for collaborators or strategic decisions (D'ippolito and Rüling, 2019; Iglič et al., 2017; Kumar, 2019) are more important for academic success and co-occur with social capital. For instance, researchers with many collaboration partners could be more successful because they managed to acquire the necessary funding for expensive high impact research and thereby provided exciting research opportunities for their fellow scientists (Paul-Hus et al., 2017; Ponomariov and Boardman, 2016). Therefore, future research should take seniority and access to resources and their interaction with social capital into account.

Despite these limitations, our paper contributes a new perspective on scientific success by including the possibility of failure. In addition, we emphasize the role of social closure depending on central vs. peripheral positions in the network, debunk the role of interdisciplinarity for scientific success and underline the importance of persistent ties. Our findings regarding the importance of persistent collaborations present a strong argument in favor of collaboration and against unlimited competition as well as short-term working contracts in academia, as described by Rhoades (2013).

Therefore, our findings have implications for policy makers, administrators, and funding agencies organizing scientific competition as well as for scholars simply seeking to advance their careers. Our analysis reveals that both amount and type of social capital are strongly linked with publishing in prominent research outlets. In particular, persistent ties seem to be essential to target high impact journals successfully. Therefore, we recommend that funding bodies and policy makers encourage long-term research projects forging persistent collaborations, which according to our findings translates into academic success more frequently than the amount collaborations as indicated by the sheer number of co-authors per arXiv preprint.

Despite the counter-intuitive finding that interdisciplinary collaboration is detrimental for publishing in high impact journals, we do not suggest that interdisciplinary projects should receive less funding or that interdisciplinary collaboration should be terminated due to strategic concerns. In contrast, our investigation points out the obstacles interdisciplinary research has to face in gaining acceptance within closed epistemic communities and the importance to overcome (sub)disciplinary boundaries. Thereby, interdisciplinary collaborations could benefit from additional resources or organizational forms such as inter-disciplinary centers (Biancani et al., 2018). However, these centers could be established for longer periods of time in order to win a place within different specialists' discourses (e.g. the Collaborative Research Centers of the German Science Foundation). Also, we would like to highlight that our analysis excluded high impact journals dedicated to interdisciplinarity like *Science* and *Nature*. Therefore, future research should broaden the scope of the analysis and investigate under which circumstances and in which types of academic outlets interdisciplinary work receives attention and recognition.

**Appendix A: Cleaning Procedures**

Even though the authors enter the names themselves, we discovered various spellings of the same name in the raw data insofar as first names are sometimes abbreviated and sometimes not. To avoid duplicates we have reduced all first names to a single letter. In addition, we excluded all corporations, groups, and teams as authors and deleted authors collaborating only with such groups, since these often consist of hundreds of physicists. This procedure allows us to focus on direct scholarly collaboration in line with Leahy and Cain (2013), and avoid the fallacy of authors with an extremely high amount of publications each year (e.g., exceeding 100 publications per year).

**Appendix B: Overview of the high impact journals used for the construction of the dependent variable.**

To assign journals correctly to a discipline and, hence, find high impact journals of each Physics subdiscipline we use the subdisciplinary differentiation and impact factors provided by *Clarivate analysis Journal Citation Reports* (Clarivate, 2020). We sorted the journals by subdisciplines and by five-year citation rate and only considered the ten leading journals of each specialization as high impact journal. This resulted in a total number of 83 high impact journals out of 456 journals represented in the data, as seven journals were listed as high impact journals in two subdisciplines. If an arXiv preprint was published in one of those 83 journals, it is considered a successful publication.

**Table B1.** List of the high impact physics journals by subdiscipline. Doubles are listed in bold.

| Sub-discipline | Physics, Applied | Physics, Atomic and Molecular | Physics, Condensed Matter | Physics, Fluids & Plasma | Physics, Mathematical |
|---|---|---|---|---|---|
| 1 | **Nature Materials** | Journal of Physical Letters | **Nature Materials** | Annual Review of Fluid Mechanics | Computer Physics Communications |
| 2 | Nature Photonics | Npj Quantum Information | Advances in Physics | Nuclear Fusion | **Communications in Nonlinear Science and Numerical Simulation** |
| 3 | Materials Science & Engineering Reports | Progress in Nuclear Magnetic Resonance Spectroscopy | Advanced Materials | Plasma Sources Science & Technology | Journal of Computational Physics |
| 4 | **Advanced Energy Materials** | Journal of Chemical Theory and Computation | Annual Review of Condensed Matter Physics | Plasma Processes and Polymers | Applied and Computational Harmonic Analysis |
| 5 | Advanced Energy Materials | International Reviews in Physical Chemistry | **Advanced Energy Materials** | Experimental Thermal and Fluid Science | Communications in Mathematical Physics |
| 6 | Applied Physics Reviews | Physical Chemistry – Chemical Physics | Surface Science Reports | Journal of Fluid Mechanics | **Physical Review E** |
| 7 | **Nano Letters** | Journal of Molecular Liquids | **Nano Letters** | **Communications in Nonlinear Science and Numerical Simulation** | Chaos |
| 8 | Nano Energy | ChemPhysChem | **Advanced Functional Materials** | Biomicrofluids | Journal of Statistical Mechanics – Theory and Experiment |
| 9 | **Advanced Functional Materials** | Structural Dynamics | **Npj Quantum Information** | Plasma Physics and Controlled Fusion | Quantum Information Processing |
| 10 | Npj Quantum Information | Journal of Chemical Physics | Small | **Physical Review E** | Communications in Computational Physics |

**Table B1. Continued.** List of the high impact physics journals by subdiscipline. Doubles are listed in bold.

| Sub-discipline | Physics, Multidisciplinary | Physics, Nuclear | Physics, Particles and Fields | Chemical Physics |
|---|---|---|---|---|
| 1 | **Progress in Particle and Nuclear Physics** | Living Reviews in Relativity | Living Reviews in Releativity | Nature Materials |
| 2 | Annual Review of Nuclear and Particle Science | **Progress in Particle and Nuclear Physics** | **Progress in Particle and Nuclear Physics** | Advanced Materials |
| 3 | Atomic Data and Nuclear Data Tables | **Journal of High Energy Physics** | **Journal of High Energy Physics** | Advanced Energy Materials |
| 4 | **Chinese Physics C** | **European Physical Journal C** | **European Physical Journal C** | Annual Review of Phyiscal Chemistry |
| 5 | Nuclear Data Sheets | **Chinese Physics C** | **Chinese Physics C** | ACS Nano |
| 6 | **Physics Letters B** | **Physics Letters B** | Physics Letters B | Surface Science Reports |
| 7 | **Journal of Physics G-Nuclear and Particle Physics** | Journal of Cosmology and Astroparticle Physics | Journal of Cosmology and Astroparticle Physics | Nano Letters |
| 8 | Physical Review C | **Physical Review D** | **Physical Review D** | Nano Energy |
| 9 | European Physical Journal A | **Annual Review of Nuclear and Particle Science** | **Annual Review of Nuclear and Particle Science** | Journal of Photochemistry and Photobiology Reviews |
| 10 | Physical Review Accelerators and Beams | **Nuclear Physics B** | **Nuclear Physics B** | Advanced Functional Materials |

# Appendix C: Descriptive Statistics

**Table C1.** Descriptive Statistics on the independent variables included in our model.

| Variable | Observations | Mean | Standard deviation | Minimum | Maximum | skewness | kurtosis |
|---|---|---|---|---|---|---|---|
| Degree Centrality | 254,846 | 1.3076 | 1.2829 | 0.3333 | 46.83736 | 6.8474 | 110.5132 |
| Eigenvector Centrality | 254,846 | 0.0002 | 0.0059 | $-7.09*10^{16}$ | 0.6245 | 64.7218 | 5047.934 |
| Persistent Ties | 254,846 | 1.0104 | 4.0575 | 0 | 99 | 10.7468 | 152.827 |
| Subdisciplinary Collaboration | 254,846 | 0.2508 | 0.4335 | 0 | 1 | 1.1500 | 2.3225 |
| Interdisciplinary Collaboration | 254,846 | 0.1795 | 0.3838 | 0 | 1 | 1.6698 | 3.7884 |

**Table C2.** Absolute and relative frequencies of the three different types of collaboration.

|  | Frequency | Percent |
|---|---|---|
| Collaboration within a single domain of physics | 145,183 | 56.97 |
| Collaboration between different domains of physics | 63,906 | 25.08 |
| Interdisciplinary collaboration | 45,757 | 17.95 |
| Σ | 254,846 | 100.00 |

**Appendix D: Robustness Checks**

To examine if our models are robust to a more broadly defined dependent variable of publication success, we use arXiv preprint published in *any* journal as first robustness check. We do that for three reasons: (1) to investigate if getting a preprint published in a high impact journal follows similar rules as getting a preprint published at all. (2) It can also be considered a success to get a preprint published in any journal. (3) It can be argued that lower ranked journals are leaning more towards interdisciplinary research, following Rafols et al. (2012). For these reasons, we compare the full multilevel event history model with getting an arXiv preprint published in a high impact journal as dependent variable against a model with getting an arXiv preprint published in any journal (see table D1).

At first glance, we see that the main effects associated with persistent ties, subdisciplinary and interdisciplinary collaboration remain robust. However, we witness some important changes in effect strength and direction of effects for degree centrality, eigenvector centrality and some of the interaction effects. Starting with degree centrality, we see no associations with the chances of getting an arXiv preprint published in any journal. At the same time, it appears that having higher volumes of social capital expressed as eigenvector centrality decreases the chances of getting published in any journal by 6.65% ($p < 0.001$) for each higher percentile. These values indicate that collaborating with less eminent colleagues might increase publication chances for non-high impact journals.

Turning to the interaction terms, we see that having persistent ties at one's disposal decreases the chances of getting a preprint published in contrast for getting an arXiv preprint published in a high impact journal. At the same time, the positive and significant interaction between interdisciplinary cooperation and persistent ties loose its significance. These findings indicate that long term interdisciplinary collaboration or long term subdisciplinary collaboration lowers the chances of getting published even further comparted to getting published in high impact journals. This is possibly due to differences in epistemic cultures.

Table D1: Exponentiated coefficients; *t* statistics in parentheses. Percentiles are used to calculate the effects of degree centrality and eigenvector centrality on the dependent variable.
$^{*} p < 0.05$, $^{**} p < 0.01$, $^{***} p < 0.001$

| Dependent variable: | Chance to publish an arXiv preprint in a high impact journal | Chance to publish an arXiv preprint in any journal |
|---|---|---|
| Degree Centrality | 0.9189*** | 1.0007 |
|  | (-14.19) | (0.47) |
| Eigenvector Centrality | 1.3859*** | 0.9335*** |
|  | (51.52) | (-50.32) |
| Persistent Ties: yes | 1.1727*** | 1.1860*** |
|  | (15.75) | (13.62) |
| Persistent collaborations: yes * Eigenvector Centrality Categorized | 0.9668*** | 0.9813*** |
|  | (-66.07) | (-10.03) |
| Persistent Ties: yes * Degree Centrality | 1.0088*** | 1.0229*** |
|  | (17.14) | (11.17) |
| Subdisciplinary Collaboration | 0.3828*** | 0.6759*** |
|  | (-37.63) | (-30.58) |
| Interdisciplinary Collaboration | 0.9232*** | 0.9199*** |
|  | (-3.22) | (-6.04) |
| Subdisciplinary Collaboration * Degree Centrality | 1.0347*** | 1.0246*** |
|  | (9.88) | (11.05) |
| Subdisciplinary Collaboration Persistent Ties: yes | 1.1101*** | 0.9931*** |
|  | (5.45) | (-7.40) |
| Subdisciplinary Collaboration * Eigenvector Centrality | 1.0498*** | 1.0201*** |
|  | (13.29) | (9.71) |
| Interdisciplinary Collaboration * Degree Centrality | 1.0029 | 1.0103*** |
|  | (0.83) | (4.34) |
| Interdisciplinary Collaboration * Persistent Ties: yes | 1.1204*** | 1.0013 |
|  | (5.90) | (0.99) |
| Interdisciplinary Collaboration * Eigenvector Centrality | 0.9807*** | 0.9897*** |
|  | (-5.36) | (-4.64) |
| N | 245432 | 245432 |
| AIC | 315470.3 | 404208.03 |
| Log lik. | -157719.2 | -202088.02 |

Simultaneously, the positive and highly significant values of interdisciplinary collaboration and degree centrality indicates that having many, occasional collaborators from different disciplines increases the chances for an arXiv preprint to get published slightly. In combination with the interaction between interdisciplinary collaborations and eigenvector centrality, these finding suggest that interdisciplinary collaboration among more peripheral scholars yield higher rates of success for getting published than interdisciplinarity among eminent scholars. Taken together, our first robustness check reveals additional obstacles for getting an arXiv preprint published in scholarly outlets.

We performed linear mixed effects regression models as second robustness check. We did so to account for unobserved variable bias introduced, for instance, by institutional effects or gender effects that could not be accounted for in our main analysis. Analogous to the first robustness check, we calculated the full model as seen in table 2 of the main document and calculated the effects of social capital and interdisciplinarity on either getting an arXiv preprint published in a high impact journal (left column in table D2), or in any journal (right column in table D2).

Differences in a number of variables appear in the two models calculated compared to the multilevel event history models, however, having persistent ties remains positive and is again among the strongest effects. Also, spanning subdisciplinary boundaries is strongly negative in both model setups. In contrast to the event history model though, there is a positive effect of interdisciplinary collaboration on getting an arXiv preprint published in high impact journals. In addition, positive interaction effects of having persistent ties and inter- or subdisciplinary collaboration vanish.

These changes may appear because a mixed effects model does not take time from the initial upload of a draft to its possible publication into account. However, the panel regression "fixes" individual attributes. Therefore, it highlights the necessity for future studies to control for effects associated with academic institutions or gender. However, the main effects of persistent ties and negative effects of research spanning (sub)disciplinary boundaries remain stable across the different statistical setups.

**Table D2:** Two-Level Mixed Effects Regression; *t* statistics in parentheses. Percentiles are used to calculate the effects of degree centrality and eigenvector centrality on the dependent variable.
* *p* < 0.05, ** *p* < 0.01, *** *p* < 0.001

| Dependent variable: | Chance to publish an arXiv preprint in a high impact journal | Chance to publish an arXiv preprint in any journal |
|---|---|---|
| Degree Centrality | 0.0006 | 0.0016** |
| | (1.26) | (3.06) |
| Eigenvector Centrality | 0.0150*** | 0.0071*** |
| | (29.36) | (13.58) |
| Persistent Ties: yes | 0.0455*** | 0.0367*** |
| | (9.80) | (7.70) |
| Persistent collaborations: yes * Eigenvector Centrality Categorized | -0.0035*** | -0.0022** |
| | (-4.58) | (-2.79) |
| Persistent Ties: yes * Degree Centrality | 0.0037*** | 0.0035*** |
| | (4.97) | (4.68) |
| Subdisciplinary Collaboration | -0.1113*** | -0.0647*** |
| | (-25.11) | (-14.15) |
| Interdisciplinary Collaboration | 0.0655*** | 0.1136*** |
| | (12.86) | (21.86) |
| Subdisciplinary Collaboration * Degree Centrality | 0.0082*** | 0.0093*** |
| | (10.42) | (11.46) |
| Subdisciplinary Collaboration Persistent Ties: yes | -0.0002 | 0.0023*** |
| | (-0.33) | (3.87) |
| Subdisciplinary Collaboration * Eigenvector Centrality | 0.0039*** | 0.0004*** |
| | (4.81) | (5.26) |
| Interdisciplinary Collaboration * Degree Centrality | -0.0017* | 0.0008 |
| | (-1.96) | (0.87) |
| Interdisciplinary Collaboration * Persistent Ties: yes | 0.0003 | 0.0005 |
| | (0.39) | (0.71) |
| Interdisciplinary Collaboration * Eigenvector Centrality | -0.0061*** | -0.0078*** |
| | (-6.64) | (-8.34) |
| *N* | 245432 | 245432 |
| *AIC* | 334,311.13 | 319,198.98 |
| ICC | 0.2221 | 0.2430 |
| Log lik. | -159,583.49 | -167,139.57 |